\documentclass[prb,twocolumn,showpacs,amsmath,amssymb]{revtex4}
\usepackage[dvips]{graphicx}
\usepackage{bm}

\begin{document}
\title{Full Counting Statistics for Number of Electrons Dwelling in a Quantum Dot}
\author{Yasuhiro~Utsumi$^{1,2}$}
\address{
$^1$Condensed Matter Theory Laboratory, RIKEN, Wako, Saitama 351-0198, Japan \\
$^2$Institut f\"{u}r Theoretische Festk\"{o}perphysik, Universit\"{a}t Karlsruhe, 76128 Karlsruhe, Germany}
\pacs{73.23.Hk,72.70.+m}

\begin{abstract}
Motivated by recent real-time electron counting experiments, we evaluate the full counting statistics (FCS) for the probability distribution of the electron number inside a quantum dot which is weakly coupled to source and drain leads. 
A non-Gaussian exponential distribution appears when there is no dot state close to the lead chemical potentials. 
We propose the measurement of the joint probability distribution of current and electron number, which reveals correlations between the two observables. 
We also show that for increasing strength of tunneling, the quantum fluctuations qualitatively change the probability distribution of the electron number.
In this paper, we derive the cumulant generating functions (CGFs) of the joint probability distribution for several cases. 
The Keldysh generating functional approach is adopted to obtain the CGFs for the resonant-level model and for the single-electron transistor in the intermediate conductance regime. 
The general form for the CGF of the joint probability distribution is provided within the Markov approximation in an extension of the master equation approach 
[D. A. Bagrets, and Yu. V. Nazarov, Phys. Rev. B {\bf 67}, 085316~(2003)].
\end{abstract}

\date{\today}
\maketitle

\newcommand{\rd}{d}
\newcommand{\ri}{i}
\newcommand{\mat}[1]{\mbox{\boldmath$#1$}}
\newcommand{\mtau}{\mbox{\boldmath$\tau$}}
\newcommand{\cgf}{{\cal W}}
\newcommand{\cf}{{\cal Z}}

\section{Introduction}

Measurements of the average current and its fluctuation (noise) 
have been powerful tools to study 
the quantum transport in mesoscopic systems~\cite{Qua-Noi-Mes-Phy}. 
In the last decade, the theory of full counting statistics (FCS)~\cite{Levitov} has been established. 
The FCS provides a probability distribution of the current $P(I)$ 
from which one obtains not only first and second cumulants, i.e. 
the average current and the noise, but also any order cumulant. 
Precisely, $I \equiv q/t_0$ is 
the time-averaged current during the measurement time $t_0$, 
where 
\begin{eqnarray}
q \! = \! \int_{-t_0/2}^{t_0/2} \!\!\!\! d t \; I(t)/e,
\label{eqn:q}
\end{eqnarray}
is the number of transmitted electrons.

\begin{figure}[ht]
\includegraphics[width=0.6 \columnwidth]{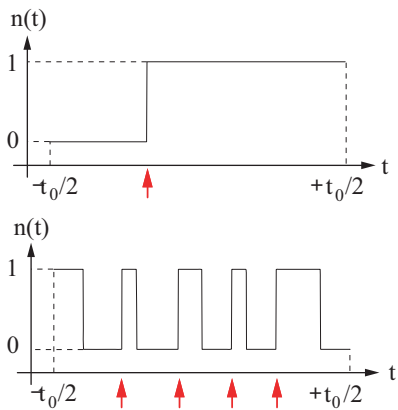}
\caption{The time evolution of electron number for rare (upper panel) and frequent (lower panel) tunneling cases.}
\label{fig:1}
\end{figure}

The last few years, there have been remarkable advances in the experimental study of the FCS~\cite{Reulet,Bomze,Gustavsson,Fujisawa}. 
The third cumulant of the probability distribution of current has been measured for a tunnel junction~\cite{Reulet,Bomze}. 
Recently, the probability distribution itself was experimentally obtained for a quantum dot (QD) coupled to source (R) and drain (L) leads~\cite{Gustavsson,Fujisawa}. 
The experiments adopted the real-time electron-counting technique using quantum point contact (QPC) charge detectors. 
The measured quantity is the time evolution of electron number inside the QD, $n(t)$. 
The experiment~\cite{Gustavsson} demonstrated that $n(t)$ fluctuates between 0 and 1, as schematically shown in Fig.~\ref{fig:1}. 
The number of transmitted electrons through the QD, $q$, is obtained by counting the number of transitions from 0 to 1 [kinks pointed by arrows in Fig.~\ref{fig:1}]. 
However, the technique works when electrons tunnel in one direction only, in which case there is a one-to-one correspondence between the changes of $n(t)$ from 0 to 1 and the tunneling processes from one lead L into the QD. 
Therefore, the analysis of the experiment~\cite{Gustavsson} has been limited to large source-drain voltages 
$T \! \ll \! eV$ (we use $k_{\rm B} \!=\! \hbar \!=\! 1$) 
where the thermal fluctuations are suppressed. 

The experiments were also performed in the weak tunneling case, 
where the FCS theory within the Markov approximation~\cite{mc,Bagrets1,Qua-Noi-Mes-Phy} works perfectly. 
For increasing strength of the tunnel coupling, the FCS is 
particularly interesting because quantum coherence effects show up. 
However, the method used in the experiment~\cite{Gustavsson} 
does not work for this case either, 
since electrons tunnel back and forth coherently. 
Therefore it is important to extend the FCS theory 
and to relate it directly with 
quantities that are measurable in the QPC experiments. 
In this paper, we will provide the FCS for the time-averaged electron number $N \equiv \tau/t_0$, where
\begin{eqnarray}
\tau \! = \! \int_{-t_0/2}^{t_0/2} \!\!\!\! d t \; n(t),
\label{eqn:n}
\end{eqnarray}
can be obtained from the experimental data. 
The observable $N$ is well defined for any condition 
including small source-drain voltages $T \! \gg \! eV$. 
On top of that, from the experimental setup in Ref.~\cite{Gustavsson}, 
one can obtain more information, 
namely the joint probability distribution of current and electron number $P(I,N)$,
since at a large source-drain voltage $T \! \ll \! eV$, 
the data from a single measurement [Fig.~\ref{fig:1}] provides 
particular values for both of $q$ and $\tau$. 

In what follows, we will calculate the characteristic function
$\cf$ 
or the cumulant generating function (CGF) $\cgf$ for the joint probability distributions, 
\begin{eqnarray}
\cf(\lambda,\xi)
=
\sum_q
\int_{-\infty}^\infty
\!\!\!\!
d \tau
\,
P(I,N)
\, 
{\rm e}^{i \, q \, \lambda \, + \, i \, \tau \, \xi}, 
\\
\cgf(\lambda,\xi) 
=
\ln 
\cf(\lambda,\xi)
= 
\sum_{k,l}
\frac{
(\ri \lambda)^k (\ri \xi)^l
}
{k! \, l!}
\langle \! \langle \delta q^k \delta \tau^l \rangle \! \rangle.
\end{eqnarray}
The parameters $\lambda$ and $\xi$ are called the `counting fields'. 
The aim of this paper is to analyze basic features of the statistical properties of electron number inside the QD and the joint probability distribution for several relevant cases. 

%
%
%

In order to derive the CGFs for the joint probability distribution, we adopt two approaches, the Keldysh generating functional~\cite{Chou,Kamenev,Kamenev1} and the master equation approaches~\cite{Bagrets1,Qua-Noi-Mes-Phy}. 
The former has been used widely in mesoscopic physics~\cite{ref1,ref2,ref3,ref4,ref5,ref6,ref7,ref8,ref9,Utsumi,Utsumi1} independently of the FCS study.
However, they are closely related to each other: 
The Keldysh generating functional is reduced to the CGF by fixing the `quantum' components of the source fields as the counting fields~\cite{Kamenev}. 
In sections~\ref{subsec:rlm} and \ref{subsec:set}, we extend our previous Keldysh generating functional study for the current and charge noises~\cite{Utsumi} to the FCS. 
%
%
In the formal level, our extension could be consistent with the FCS theory of charge developed by Pilgram and B\"uttiker~\cite{Pilgram}. 
Their scheme is based on Levitov's $S$-matrix approach~\cite{Levitov} and 
suitable to deal with an open and noninteracting chaotic cavity, since it provides a systematic method for the disorder averaging~\cite{Pilgram}. 
At the same time, the scheme requires more involved calculations to account for the effects of the back scattering from the tunnel barriers and the Coulomb interaction. 
Therefore Ref.~\cite{Pilgram} analyzes the situations where both of them are weak. 
The present paper will address complementary cases, closed and interacting QDs. 
In Sec.~\ref{subsec:rlm}, we consider a simple model for a non-interacting QD, the resonant-level model (RLM). 
Through the calculations we would like to demonstrate that the Keldysh generating functional is a useful tool for the FCS study. 
In Sec.~\ref{subsec:set}, we consider an interacting QD, 
the single-electrons transistor (SET) in an intermediate conductance regime. 
There the approximate generating functional derived in Refs.~\cite{Utsumi,Utsumi1} is extended to the CGF. 
In Sec.~\ref{subsec:mastereq}, we also derive the general CGF for the joint probability distribution of current and electron number within the Markov approximation: We extend the master equation approach originally developed for the probability distribution of current only by Bagrets and Nazarov~\cite{Bagrets1,Qua-Noi-Mes-Phy}.

%
%

The structure of the paper is the following. 
In Sec.~\ref{sec:1}, we discuss the probability distribution of electron number for the spinless non-interacting QD, the RLM. 
Then in Sec.~\ref{sec:2}, the spin effect in an interacting QD is considered within the Markov approximation. 
In Sec.~\ref{sec:3} we present the joint probability distribution of current and electron number. 
In Sec.~\ref{sec:4}, we will demonstrate how the quantum fluctuations of charge affect on the statistical properties of electron number inside the QD.
We summarize in Sec.~\ref{sec:5}.

\section{FCS for a non-interacting QD}
\label{sec:1}

In this section, we derive the CGF for the RLM.
Our calculational tool is the Keldysh generating functional in the path-integral representation 
(We refer to Refs.~\cite{Kamenev,Kamenev1} for details). 
We will check that our result is consistent with previous theories up to the second cumulant (Sec.~\ref{subsec:rlm}). 
In Sec.~\ref{subsec:expdist}, we discuss the equilibrium probability distribution of electron number in the weak tunneling regime.

\subsection{FCS formalism based on the Keldysh generating functional}
\label{subsec:rlm}

The Hamiltonian of a non-interacting closed QD for spinless case (RLM) is given by, 
\begin{equation}
\hat{H}
=
\sum_{{r} k} 
\varepsilon_{k} 
\hat{a}_{{r} k}^\dagger 
\hat{a}_{{r} k}
+
\epsilon_0 \, 
\hat{d}^\dagger \hat{d}
+
\sum_{{r} k} 
(
V_{r} \hat{d}^\dagger \hat{a}_{{r} k} + {\rm H.c.}
)
\, ,
\label{eqn:H}
\end{equation}
where $\hat{a}_{r k}$ annihilates an electron with wave vector $k$ in the lead $r$ while, $\hat{d}$ annihilates an electron in the QD with the dot-level $\epsilon_0$. 
The electrons in the lead $r$ obey a Fermi distribution 
$f(\omega - \mu_{r}) = 1/({\rm e}^{(\omega - \mu_{r})/T}+1)$.
The chemical potential $\mu_{r}$ is fixed by the source-drain voltage
$eV \!=\! \mu_{L} \!-\! \mu_{R}$.

\begin{figure}[ht]
\includegraphics[width=.6 \columnwidth]{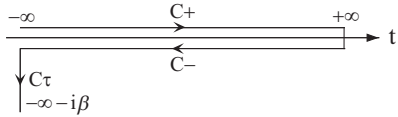}
\caption{
The closed time-path consisting of forward $C_{+}$, backward $C_{-}$ and imaginary time $C_{\tau}$ branches.
}
\label{fig:CTP}
\end{figure}

In order to derive the CGF for the joint probability distribution of $q$ and $\tau$, equivalently $I$ and $N$, we first introduce the closed time-path $C$ (Fig.~\ref{fig:CTP}). 
The choice of this time-path is particularly convenient since with the help of the technical know-how to perform the path integral~\cite{Kamenev,Kamenev1} and to solve the differential equation for the Green function (GF) defined on $C$~\cite{Niemi}, such as Eq.~(\ref{eqn:gf_eqn}), the calculations of the characteristic function become quite similar to those of the partition function in the imaginary-time formalism.
Next, we introduce two source fields, 
$\varphi_{r}(t)$ and $h(t)$, the phase of the tunneling matrix element and 
the fluctuation of the dot-level~\cite{Utsumi}:
\begin{eqnarray}
V_{r} \rightarrow V_{r} e^{i \varphi_{r}(t)},
\;\;\;\;
\epsilon_0 \rightarrow \epsilon_0 - h(t). 
\end{eqnarray}
%
Note that here the time $t$ is defined on the closed time-path $C$. 
After project the time on $C$ onto the real axis, two components, 
$\varphi_{{r} +}$($h_+$) and $\varphi_{{r} -}$($h_-$) residing $C_+$ and $C_-$, appear. 
The averages of the two components, 
\begin{eqnarray}
\frac{\varphi_{{r} +}(t)+\varphi_{{r} -}(t)}{2}=\mu_{r} t,
\;\;\;\;
\frac{h_+(t)+h_-(t)}{2}=0,
\label{eqn:so_ph}
\end{eqnarray}
posses the physical meaning. 
The differences are fictitious degrees of freedom called the `quantum' components~\cite{Kamenev}. 
In the generating functional approach, these components are set to 0 in the end of calculations~\cite{Chou,Kamenev,Kamenev1}. 
In the FCS scheme, they are fixed as constants, the counting fields, 
\begin{eqnarray}
\varphi_{{r} +}(t)-\varphi_{{r} -}(t)=\lambda_r,
\;\;\;\;
h_+(t)-h_-(t)=\xi,
\label{eqn:so_qu}
\end{eqnarray}
during the measurement ($-t_0/2 \!<\! t \!<\! t_0/2$) and 0 otherwise. 

We introduce the path-integral representation of the Keldysh generating functional from the Hamiltonian (\ref{eqn:H}) by the time-slicing method~\cite{Kamenev,Kamenev1}. 
Under the condition (\ref{eqn:so_qu}), it is reduced to the CGF: 
\begin{eqnarray}
\cgf(\lambda,\xi) 
\! = \!
\ln \int
\!\!
{\cal D} 
[a_{{r} k}^*,  a_{{r} k},  d^*, d ]
\,
\exp 
\left(  
i S
\right),
\end{eqnarray}
\begin{eqnarray}
S
&=&
\int_C d t
\left \{
\sum_{{r} k}
a_{{r} k}(t)^* 
(i \, \partial_t-\varepsilon_k) 
\, 
a_{{r} k}(t)
\right.
\nonumber \\
&+&
d(t)^* [i \, \partial_t-\epsilon_0+h(t)] \,d(t)
\nonumber \\
&-&
\left.
\sum_{{r} k}
(
V_{r} \,
{\rm e}^{\ri \varphi_{r}(t)} \,
d(t)^*
a_{{r} k}(t)
+{\rm c. c.}
)
\right \},
\label{eqn:action_RLM}
\end{eqnarray}
where Grassmann fields satisfy the antiperiodic boundary condition,
for example 
$d(-\infty \!-\! i \beta) \! = \! -d(-\infty)$.
Although the counting fields $\lambda_{L}$ and $\lambda_{R}$ appear at this stage, the final result contains only 
$\lambda \! \equiv \! \lambda_{L} \! - \! \lambda_{R}$ 
from the charge conservation \cite{Belzig}. 
We can check the first derivative in $\xi$ gives $\tau$ with an offset;
\begin{eqnarray}
\left.
\frac{\partial \cgf}{\partial (i \xi)} 
\right |_{\lambda=\xi=0} =
\frac{1}{2}
\int^{t_0/2}_{-t_0/2}
\!\!\!\!
d t \, 
\langle [ \hat{d}^\dagger(t), \hat{d}(t)] \rangle
=
\tau-\frac{t_0}{2},
\label{eqn:average}
\end{eqnarray}
where $\hat{d}(t)$ is in the Heisenberg picture and the average is performed over the Hamiltonian (\ref{eqn:H}) without the tunneling term
(For technical details, see Appendix.~A of Ref.~\cite{Utsumi}). 
From now on, we will shift $N$ by the offset $-1/2$.

After the path integral over $a_{rk}$ along $C$~\cite{Kamenev,Kamenev1}, 
which is formally to complete the square, 
we obtain the effective action for the QD as (In the following, we will omit trivial constants), 
\begin{eqnarray}
S_d
=
\int_C 
\!\!
d t \, d t' \,
d(t)^* 
\left[
\,
g^{-1}(t,t')
-
\Sigma^{\lambda,\xi}(t,t')
\,
\right]
d(t').
\label{eqn:d_action}
\end{eqnarray}
The GF satisfies the following equation and the anti-periodic boundary condition: 
\begin{eqnarray}
(i \, \partial_t-\epsilon_0) \, g (t,t')
=
\delta (t,t'),
\label{eqn:gf_eqn}
\\
g(-\infty + i \beta,t')=-g(-\infty,t'). 
\label{eqn:antiperiodic}
\end{eqnarray}
The $\delta$-function is defined on $C$ and satisfies 
$\int_C dt \, \delta(t,t')=1$. 
Precisely $g$ is a discrete matrix, since time is discrete~\cite{Kamenev,Kamenev1}. 
In the continuous notation, we can regard Eq.~(\ref{eqn:gf_eqn}) as the differential equation defined on $C$, which is given as a function $t=z(s)$ with $s$, a real and monotonically increasing parameter. 
Then Eq.~(\ref{eqn:gf_eqn}) can be solved~\cite{Niemi,Utsumi1} by considering the anzatz~\cite{Niemi},
\begin{eqnarray}
g(t,t')=g^{-+}(t-t') \, \theta(t,t')+g^{+-}(t-t') \, \theta(t',t),
\label{eqn:anzatz}
\end{eqnarray}
where the step function is defined on $C$: 
$
\partial_t \, \theta(t,t')=-\partial_{t' } \, \theta(t,t')=\delta(t,t'). 
$
We observe that $g^{\pm \mp}$ becomes, 
\begin{eqnarray}
g^{-+}(t \!-\! t') \!=\! (1\!-\!a) \, {\rm e}^{-i \epsilon_0 (t \!-\! t')},
\;\;
g^{+-}(t \!-\! t') \!=\! - a \, {\rm e}^{-i \epsilon_0 (t \!-\! t')}. 
\nonumber
\end{eqnarray}
The anti-periodic boundary condition (\ref{eqn:antiperiodic}), 
$g^{-+}(-\infty-i \beta,t')=-g^{+-}(-\infty,t')$, 
determines the parameter $a$ as $a=f(\epsilon_0)$. 
After projecting the time on $C$ onto the real axis, Eq.~(\ref{eqn:anzatz}) is mapped onto the $2 \times 2$ Keldysh space:
\begin{eqnarray}
\hat{g}(t,t')
&=&
\left[
\begin{array}{cc}
g^{++}(t \!-\! t')
&
g^{+-}(t \!-\! t')
\\
g^{-+}(t \!-\! t')
&
g^{--}(t \!-\! t')
\end{array}
\right],
\nonumber \\
g^{++}(t \!-\! t')
\! &=& \!
g^{+-}(t \!-\! t') \, \theta(t \!-\! t') \!+\! g^{-+}(t \!-\! t') \, \theta(t' \!-\! t),
\nonumber \\
g^{--}(t \!-\! t')
\! &=& \!
g^{+-}(t \!-\! t') \, \theta(t' \!-\! t) \!+\! g^{-+}(t \!-\! t') \, \theta(t \!-\! t').
\nonumber
\end{eqnarray}
A compact representation follows from the Keldysh rotation 
by an orthogonal matrix 
$Q \! = \! (\mtau^0 \! - \! i \, \mtau^2)/\sqrt{2}$
as 
$\tilde{g} \!=\! Q \, \hat{g} \, Q^{-1}$
[
$\mtau^i$ ($i \! = \! 1,2,3$) are the Pauli matrices and $\mtau^0$ is the unit matrix
].
Then in Fourier space, the matrix GF is written with the retarded, 
$g^R(\omega) \!=\! 1/(\omega+ i 0-\epsilon_0)$, 
the advanced 
$g^A \!=\! g^{R \, *}$
and the Keldysh
$g^K(\omega)=2 i \, {\rm Im} \, g^R(\omega) \tanh(\omega/2T)$
components as, 
\begin{equation}
\tilde{g}(\omega)
=
\left[
\begin{array}{cc}
0           & g^A(\omega) \\
g^R(\omega) & g^K(\omega)
\end{array}
\right]. 
\nonumber
\end{equation}
%
The self-energy in Eq.~(\ref{eqn:d_action}) contains the source fields: 
\begin{eqnarray}
\Sigma^{\lambda,\xi}(t,t')
=
\!\!\!\!
\sum_{r=L,R}
\!\!
\Sigma_r(t,t')
\, {\rm e}^{i \, (\varphi_r(t)-\varphi_r(t'))}
\!-\! 
h(t) \, \delta(t,t').
\nonumber 
\end{eqnarray}
The `bare' part 
$
\Sigma_r(t,t')
=
\!\!
\sum_k |V_{rk}|^2
\,
g_{rk}(t,t')
$
is written with the lead-electron GF $g_{rk}$, which is obtained from $g$ by replacing $\epsilon_0$ with $\varepsilon_{{r} k}$. 
The retarded and Keldysh components of $\tilde{\Sigma}_{r}$, 
\begin{eqnarray}
\Sigma_{r}^R
&=&
-i 
\frac{\Gamma_{r}}{2},
\label{eqn:reseenrlm}
\\
\Sigma_{r}^K
&=&
-i 
\Gamma_{r} \, \tanh \frac{\omega}{2 T},
\label{eqn:keseenrlm}
\end{eqnarray}
are then written with the coupling strength 
$
\Gamma_{r}
\! = \! 
2 \pi \sum_k |V_{r}|^2 \delta(\omega - \varepsilon_{k})
$, 
which is assumed to be energy independent. 

Further performing the path integral over $d$, we obtain, 
\begin{eqnarray}
\cgf
=
{\rm Tr}
\ln \left (
i {G^{\lambda, \xi}}^{-1}
\right),
\;\;\;\;
{G^{\lambda, \xi}}^{-1}
=
g^{-1}-\Sigma^{\lambda,\xi},
\label{eqn:CGFLL}
\end{eqnarray}
where the trace is understood as the time integrations along $C$. 
It can be performed utilizing the asymptotic form, 
\begin{eqnarray}
\partial_\xi 
\cgf
&=&
\int_C
\!\!\!\
\!\!
dt 
\,
G^{\lambda,\xi}(t,t) \, 
\partial_\xi 
h(t)
\nonumber \\
&=&
\frac{1}{2}
\int_{-t_0/2}^{t_0/2}
\!\!\!\!\!\!
dt
\,
{\rm Tr}
\left[
\tilde{G}^{\lambda,\xi}(t,t)
\right]
\sim
\frac{t_0}{2}
{\rm Tr}
\left[
\tilde{G}^{\lambda,\xi}(0)
\right],
\nonumber
\end{eqnarray}
for $t_0 \! \rightarrow \! \infty$. 
After the Fourier transform and the integration in terms of $\xi$, 
we obtain, 
\begin{eqnarray}
\cgf
&=&
\frac{t_0}{2 \pi}
\!\!
\int d \omega 
\,
{\rm Tr}
\ln 
\tilde{G}^{\lambda,\xi}(\omega)^{-1}, 
\nonumber
\\
\tilde{G}^{\lambda,\xi}(\omega)^{-1}
&=&
(\omega-\epsilon_0) \, \mtau_1
\!
+
\frac{\xi}{2}
\mtau_{0}
\nonumber \\
\!\!\!\!
&-&
\!\!\!\!
\sum_{r=L,R} 
\!\!
{\rm e}^{i \lambda_{r} \mtau_1/2}
\mtau_1
\tilde{\Sigma}_{r}(\omega) 
\mtau_1
{\rm e}^{-i \lambda_{r} \mtau_1/2}.
\nonumber
\end{eqnarray}
The frequency in $\tilde{\Sigma}_{r}(\omega)$ is shifted as 
$\omega \rightarrow \omega-\mu_r$ 
in order to absorb the part $\mu_r t $ in the phase, Eq.~(\ref{eqn:so_ph}). 
After straightforward calculations and 
subtracting a constant in order to fulfill the normalization condition $\cgf(0,0) = 0$, we come to the final form: 
\begin{eqnarray}
\cgf
&=&
\frac{t_0}{2 \pi} 
\! \int \!\! d \omega
\ln
\biggl \{
1
\nonumber \\
&+& \,
{\cal T} (\omega) 
f(\omega-\mu_{L}) 
[1-f(\omega-\mu_{R})]
(e^{i \lambda} - 1)
\nonumber \\
&+& \,
{\cal T} (\omega) 
f(\omega-\mu_{R}) 
[1-f(\omega-\mu_{L})]
(e^{-i \lambda} - 1)
\nonumber \\
&-&
|G^R(\omega)|^2
\biggl(
\frac{\xi^2}{4}
-
\frac{\xi}{2}
\sum_{r=L,R} \Sigma_{r}^K(\omega)
\biggl)
\biggl \}, 
\label{eqn:CGF}
\\
{\cal T} (\omega)
&=&
4 \, |G^R(\omega)|^2 \,
{\rm Im} \Sigma_{L}^R(\omega) \,
{\rm Im} \Sigma_{R}^R(\omega),
\label{eqn:T}
\\
G^R(\omega)
&=&
\frac{1}{\omega-\epsilon_0-\sum_{r=L,R} \Sigma_{r}^R(\omega)},  
\label{eqn:GF}
\end{eqnarray}
where the transmission probability 
${\cal T} (\omega)
=
\Gamma_{L} \Gamma_{R}/
[(\omega-\epsilon_0)^2+\Gamma^2/4]$
is the Lorentzian ($\Gamma=\Gamma_{L}+\Gamma_{R}$).
The fourth line of Eq.~(\ref{eqn:CGF}) is our new result. 
Without the fourth line, Eq.~(\ref{eqn:CGF}) is reduced to the Levitov-Lesovik formula~\cite{Levitov}. 

Further, the integration in frequency can be performed for $T \gg \Gamma$ following the procedure in Sec.\ IV of Ref.~\cite{Bagrets1}, 
\begin{eqnarray}
\cgf^{(1)}(\lambda,\xi)
\! = \!
t_0 \, 
\Gamma_\Sigma
\frac{\sqrt{D}-1}
{2},
\;\;\;\;
\Gamma_\Sigma=\Gamma^++\Gamma^-,
\label{eqn:cgf1st}
\\
D
=
1+
4 \frac{\Gamma^+_{L} \Gamma^-_{R}}
{\Gamma_\Sigma^2}
\, ({\rm e}^{i \lambda}\!-\!1)
+
4 \frac{\Gamma^+_{R} \Gamma^-_{L}}
{\Gamma_\Sigma^2}
\, ({\rm e}^{-i \lambda}\!-\!1)
\nonumber \\
+
\frac{
2 \, \ri \, \xi \, 
(\Gamma^+ \!-\! \Gamma^-)-\xi^2
}{\Gamma_\Sigma^2},
\nonumber
\end{eqnarray}
where $\Gamma^\pm = \sum_{r = L,R} \Gamma^\pm_{r}$ 
and 
\begin{eqnarray}
\Gamma_{r}^\pm
=
\Gamma_{r}
\,
f(\pm \epsilon_0 \mp \mu_{r}),
\label{eqn:tunnelingratelrm}
\end{eqnarray}
is the tunneling rate of an electron into/out of the QD through the junction ${r}$ within Fermi's golden rule. 
Note that the condition $T \! \gg \! \Gamma$ is that for the Markov approximation and that Eq.~(\ref{eqn:cgf1st}) is 
the first order expansion of Eq.~(\ref{eqn:CGF}) in $\Gamma$ assuming $\xi \! \propto \! \Gamma$. 
Actually, the first and second cumulants, 
$\langle\!\langle \delta \tau \rangle\!\rangle$ and 
$\langle\!\langle \delta \tau^2 \rangle\!\rangle$,
and the covariance 
$\langle\!\langle \delta q \, \delta \tau \rangle\!\rangle$, 
\begin{eqnarray}
\langle\!\langle \delta \tau \rangle\!\rangle/t_0
=
\frac{\Gamma^+ \! - \! \Gamma^-}{2 \Gamma_\Sigma},
\label{eqn:number_mean}
\\
\langle\!\langle \delta \tau^2 \rangle\!\rangle/t_0
=
2 \frac{\Gamma^+ \Gamma^-}{\Gamma_\Sigma^3},
\label{eqn:number_variance}
\\
\langle\!\langle \delta q \, \delta \tau \rangle\!\rangle
=
-2 
\frac{
\langle\!\langle \delta q \rangle\!\rangle
\langle\!\langle \delta \tau \rangle\!\rangle}
{\Gamma_\Sigma \, t_0},
\end{eqnarray}
are reproduced by the master equation approach for the noise~\cite{Korotkov}. 
Moreover, later in Sec.~\ref{subsec:mastereq}, we will rederive Eq.~(\ref{eqn:cgf1st}) in an extension of the master equation approach. 
The second check is to compare the expressions for the second cumulants and the covariance obtained from Eq.~(\ref{eqn:CGF}) with those obtained by the standard Keldysh diagrammatic approach for the Anderson model~\cite{Hershfield}. 
We check our results, 
\begin{eqnarray}
\langle\!\langle \delta \tau^2 \rangle\!\rangle/t_0
&=&
\Gamma^2
\int \! \frac{\rd \omega}{2\pi}
\frac{f_{\rm eff}(\omega) [1-f_{\rm eff}(\omega)]}
{[ (\omega-\epsilon_0)^2+\Gamma^2/4 ]^2},
\label{eqn:number_mean_general}
\\
\langle\!\langle \delta \tau \delta q \rangle\!\rangle/t_0
&=&
\Gamma
\int \! \frac{\rd \omega}{2\pi}
\frac{1/2-f_{\rm eff}(\omega)}
{(\omega-\epsilon_0)^2+\Gamma^2/4}
\nonumber \\
&\times&
{\cal T}(\omega)
[ f(\omega-\mu_{L}) -f(\omega-\mu_{R}) ],
\label{eqn:number_variance_general}
\\
f_{\rm eff}(\omega) 
&=& 
\sum_{r=L,R}
\Gamma_{r} \, f(\omega-\mu_{r})/\Gamma,
\nonumber
\end{eqnarray}
are consistent with those calculated by Hershfield [Eqs. (87) and (88) in Ref.~\cite{Hershfield}], except for a factor 2 and an offset. 
The factor 2 is lost since we consider the spinless case. The offset would be related with the shift appeared already in the average Eq.~(\ref{eqn:average}).

\subsection{Exponential distribution}
\label{subsec:expdist}

\begin{figure}[ht]
\includegraphics[width=0.85 \columnwidth]{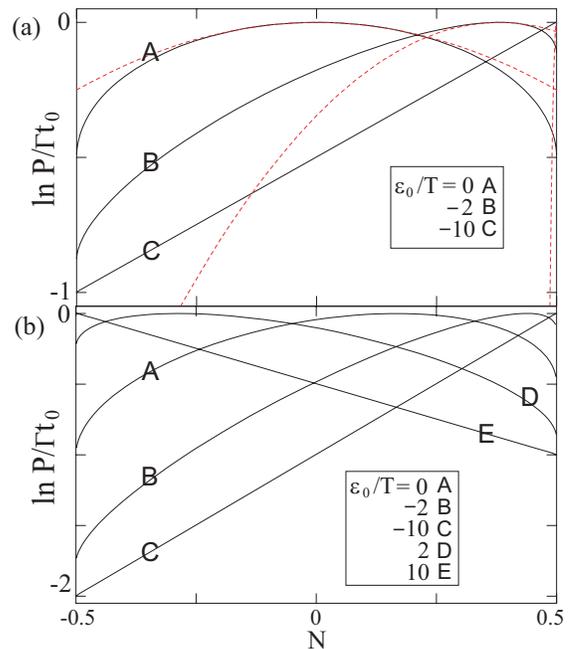}
\caption{
The logarithm of the probability distribution of electron number $P(N)$ for the noninteracting (a) and interacting (b) QDs in equilibrium for the symmetric case ($\Gamma_{L} \!=\! \Gamma_{R}$). 
The vertical axes are normalized by the coupling strength $\Gamma$ and the horizontal axes are shifted by $-1/2$. 
Solid lines are for various dot-levels. 
Dashed lines are for Gaussian approximations. 
}
\label{fig:P_N}
\end{figure}

In the limit of $t_0 \rightarrow \infty$, the number distribution follows from the inverse Fourier transform of the characteristic function $\cf$ within the saddle point approximation:
\begin{eqnarray}
P(N)=
\frac{t_0}{2 \pi} \!
\int_{-\infty}^{\infty} 
\!\!\!\! 
d \xi \, {\rm e}^{\cgf(0,\xi)-i t_0 N \xi}
\approx
{\rm e}^{\cgf(0,\xi^*)-i t_0 N \xi^*}, 
\nonumber
\\
i t_0 \, N \! = \! \partial_\xi \cgf(0,\xi^*). 
\nonumber
\end{eqnarray}
%
Solid lines in Fig.~\ref{fig:P_N} (a) are the equilibrium probability distributions of electron number in the weak tunneling regime derived from Eq.~(\ref{eqn:cgf1st}) for the symmetric coupling ($eV \!=\! 0$ and $\Gamma_{L} \! = \! \Gamma_{R}$). 
When the dot-level is at the lead chemical potentials, $\epsilon_0 \!=\! 0$ (line A), the distribution is well fitted by a Gaussian distribution around $N \approx 0$ (dashed line), 
\begin{equation}
P(N) \approx \exp 
\left[
-
t_0
\frac{(N - \langle\!\langle \delta \tau \rangle\!\rangle/t_0)^2}
{2 \, \langle\!\langle \delta \tau^2 \rangle\!\rangle/t_0}
\right],
\end{equation}
where
the first and second cumulants are given by 
Eqs.~(\ref{eqn:number_mean}) and (\ref{eqn:number_variance}).
Around $N=\pm 1/2$, it is suppressed strongly.
Since our RLM possesses the particle-hole symmetry, the distribution appears symmetric for negative and positive $\epsilon_0$ (not shown). 

The deviation from the Gaussian distribution grows rapidly as the dot-level leaves away from the chemical potential. 
The range where the Gaussian distribution is valid shrinks (lines B and C). 
Especially, for $\mp \epsilon_0 \! \gg \! T$, the exponential-like distribution appears (line C);
\begin{equation}
P(N) \approx \exp [ \, t_0 \Gamma_\Sigma \, (\pm N-1/2)]. 
\nonumber
\end{equation}
The exponent is the decay rate of an excited state, namely for $\mp \epsilon_0 \! \gg \! T$, the decay rate of an empty/occupied state. 
%
%
%
%
The strong asymmetric shape of the exponential distribution is in clear contrast with the distribution of the open chaotic cavity~\cite{Pilgram}, which should be symmetric around $N \!=\! \langle N \rangle$ in equilibrium.

%
%

In the limit of the occupied or empty QD, a more careful treatment on the branch cut of the CGF is required. 
The square root in Eq.~(\ref{eqn:cgf1st}) reads $\sqrt{D} \approx \sqrt{(\pm 1 + i \xi/\Gamma_\Sigma)^2}$ for $ \mp \epsilon_0/T \! \rightarrow \! \infty$. 
Then the branch is chosen uniquely from the normalization condition 
$\cgf^{(1)}(0,0) \!=\! 0$. 
We obtain $\cgf^{(1)} \!=\! \pm i \, t_0 \, \xi/2$ 
and thus the delta distribution, 
\begin{equation}
P(N)=\delta(N \mp 1/2). 
\nonumber
\end{equation}
It supports our intuition that without the thermal and quantum fluctuations, 
an electron (a hole) is localized in the QD.

%
%

\section{Spin effect in an interacting QD within the Markov approximation}
\label{sec:2}

Up to now, we omitted the spin degrees of freedom. 
The trivial extension of the Hamiltonian (\ref{eqn:H}) is to introduce the spin index $\sigma = \uparrow,\downarrow$ for fermions as, 
$\hat{a}_{{r} k} \! \rightarrow \! \hat{a}_{{r} k \sigma}$
and  
$\hat{d} \! \rightarrow \! \hat{d}_\sigma$. 
The extension results in the CGF twice as large as Eq.~(\ref{eqn:CGF}) because of the spin degeneracy. 
A nontrivial spin effect appears when the on-site Coulomb interaction
$\hat{H}_{\rm int}
\!=\!
U
\hat{d}_\uparrow^\dagger
\hat{d}_\uparrow
\hat{d}_\downarrow^\dagger
\hat{d}_\downarrow$ is accounted for. 
For simplicity, we will consider the limit $U \! \rightarrow \! \infty$, 
where the double occupancy is forbidden and thus the QD states are limited to 
$|\! \uparrow \rangle$, $|\! \downarrow \rangle$ and $|0 \rangle$. 
Furthermore, we will limit ourselves to the weak tunneling case $T \! \gg \! \Gamma$, where the experiments have been done~\cite{Gustavsson,Fujisawa} and the master equation approach~\cite{Bagrets1,Qua-Noi-Mes-Phy} works. 

In this section, first we construct the FCS theory for the joint probability distribution of current and electron number within the Markov approximation following the theory by Bagrets and Nazarov~\cite{Bagrets1,Qua-Noi-Mes-Phy} (Sec.~\ref{subsec:mastereq}). 
Then we will discuss the spin effect in Sec.~\ref{subsec:mastereq_result}.

\subsection{Extension of the FCS theory in the master equation approach}
\label{subsec:mastereq}

Within the Markov approximation, a particular time evolution of the QD state is characterized by $s$ tunneling events happen via a junction $r_i$ into/out of ($v_i=\pm1$) the QD at successive times in a time interval 
($-t_{\infty} < t_1 < t_2 < \cdots < t_s < t_\infty$). 
Such a sequence of the events, the sample, is then written as, 
\begin{eqnarray}
\zeta_s =(t_1,r_1,v_1; t_2,r_2,v_2; \cdots; t_s,r_s,v_s).
\nonumber
\end{eqnarray}
In the following, we assume $t_{\infty} \! \gg \! t_0 \! \gg \! 1/\Gamma$. 
The sample $\zeta_s$ determines the time evolution of the QD state 
$
l_0
\rightarrow
l_1
\cdots
l_{s-1}
\rightarrow
l_s
$. 
It also determines the time evolutions of the current through the junction $r$ into the QD and the electron number inside the QD as, 
\begin{eqnarray}
I_r(t,\zeta_s)
&=&
e
\sum_{j=1}^s
v_j \, 
\delta(t-t_j)
\,
\delta_{r,r_j},
\label{eqn:izeta}
\\
n(t,\zeta_s)
&=&
\sum_{j=1}^s
N_{l_j}
[\,
\theta(t-t_j)
-
\theta(t-t_{j+1})
\,].
\label{eqn:nzeta}
\end{eqnarray}
Here $N_l$ is the number of electrons inside the QD for the state $l$.
The probability to find the QD in a state $m$ evolves according to the master equation, 
\begin{eqnarray}
\partial_t
\,
p_m(t)
=
-
\sum_n 
L_{mn}
\, 
p_n(t). 
\nonumber
\end{eqnarray}
The off-diagonal components of the transition matrix $L$ are given by the tunneling rate within Fermi's golden rule Eq.~(\ref{eqn:tunnelingratelrm}), 
\begin{eqnarray}
L_{mn}=- \sum_{r=L,R} \Gamma_r^{\pm}
\;
\delta_{N_m,N_n \pm 1}.
\label{eqn:ODelement}
\end{eqnarray}
The diagonal components satisfy, 
\begin{eqnarray}
L_{nn} = - \sum_{m \ne n} L_{mn} = \gamma(n). 
\label{eqn:Delement}
\end{eqnarray}
%
The probability for the sample $\zeta_s$ to occur 
$Q_s(\{ t_i,r_i,v_i \})
\equiv
Q(t_1,r_1,v_1; \cdots ; t_s,r_s,v_s)
$ 
is then determined as, 
\begin{eqnarray}
Q_0
&=&
{\rm e}^{-2 \, \gamma(l_0) \, t_\infty},
\nonumber \\
Q(t_1,r_1,v_1)
&=&
{\rm e}^{-\gamma(l_1) \, (t_{\infty}-t_1)} \, \Gamma^{v_1}_{r_1}
{\rm e}^{-\gamma(l_0) \, (t_1+t_{\infty})},
\nonumber \\
&\cdots&,
\nonumber \\
Q_s(\{ t_i,r_i,v_i \})
&=&
\left( 
\,
\prod_{j=1}^{s}
{\rm e}^{-\gamma(l_j) \, (t_{j+1}-t_j)} \, \Gamma^{v_j}_{r_j}
\right)
\nonumber \\
&\times&
{\rm e}^{-\gamma(l_0) \, (t_1+t_{\infty})},
\label{eqn:prob}
\end{eqnarray}
where $t_{s+1} \!=\! t_\infty$. 
In the Markov approximation, the current $I(t)$ and the number $n(t)$, Eqs. (\ref{eqn:q}) and (\ref{eqn:n}), are simply numbers and commute at different times. 
Then the characteristic function of the joint probability distribution of 
$q_{r} \! = \! \int_{-t_0/2}^{t_0/2} d t \; I_{r}(t)/e$ and $\tau$ is, 
\begin{eqnarray}
\cf(\lambda,\xi)
=
\left \langle 
O
\right \rangle_\Omega,
\;
O=
\exp
\left (
\sum_r
i \lambda_r \, q_r
+
i \, \xi \, \tau
\right ). 
\label{eqn:chara}
\end{eqnarray}
Here the average is performed over all samples, 
\begin{eqnarray}
\left \langle
O
\right \rangle_\Omega
&=&
O_0 \, 
Q_0
+
\!\!\!\!
\sum_{ r_1,v_1 }
\int_{-t_\infty}^{t_\infty}
\!\!\!\!
\!\!\!\!
d t_1 
O(t_1,r_1,v_1)
\,
Q(t_1,r_1,v_1)
\nonumber \\
\!\!\!\!
&+&
\!\!\!\!
\sum_{ \{ r_i,v_i \} }
\!
\int_{-t_\infty}^{t_\infty}
\!\!\!\!
\!\!\!\!
d t_2
\int_{-t_\infty}^{t_2}
\!\!\!\!
d t_1
O_2(\{ t_i,r_i,v_i \})
\,
Q_2(\{ t_i,r_i,v_i \})
\nonumber \\
&+&
\cdots. 
\nonumber
\end{eqnarray}
%
For the sample $\zeta_s$, the stochastic variables $q_{r}$ and $\tau$ read, 
$
q_{{r} \, s}(\{ t_i,r_i,v_i \})
=
\sum_{j=j_{\rm min}}^{j_{\rm max}}
v_j
\,
\delta_{r,r_j}
$
and 
$
\tau_s(\{ t_i,r_i,v_i \})
=
\sum_{j=j_{\rm min}}^{j_{\rm max}}
N_{l_j} \, (t_{j+1}-t_j)
+
O(1/\Gamma)$
from Eqs.~(\ref{eqn:izeta}) and (\ref{eqn:nzeta}) and $j_{\rm min(max)}$ is the minimum (maximum) $j$ satisfying the condition $|t_j| < t_0/2$. 
The stochastic variable $O$ then reads, 
\begin{eqnarray}
O_s(\{ t_i,r_i,v_i \})
\nonumber \\
=
\!\!
\prod_{j=j_{\rm min}}^{j_{\rm max}}
\!\!
\exp 
\left(
\sum_r i \, \lambda_r \, v_j \, \delta_{r, r_j}
+
i \, \xi \, N_{l_j} \, (t_{j+1}-t_j)
\right),
\label{eqn:o}
\end{eqnarray}
for $t_0 \! \gg \! 1/\Gamma$. 
By comparing Eq.~(\ref{eqn:prob}) and Eq.~(\ref{eqn:o}), we observe that the characteristic function (\ref{eqn:chara}) can be expressed in a simple form as,  
\begin{eqnarray}
\left \langle
O
\right \rangle_\Omega
&=&
Q_0^{\lambda,\xi}
+
\sum_{ r_1,v_1 }
\int_{-t_\infty}^{t_\infty}
\!\!\!\!
d t_1
\; 
Q^{\lambda,\xi}(t_1,r_1,v_1)
\nonumber \\
&+&
\sum_{ \{ r_i,v_i \} }
\int_{-t_\infty}^{t_\infty}
\!\!\!\!
d t_2
\int_{-t_\infty}^{t_2}
\!\!\!\!
d t_1
\; 
Q_2^{\lambda,\xi}(\{ t_i,r_i,v_i \})
+
\cdots, 
\nonumber \\
\label{eqn:Dyson}
\end{eqnarray}
where 
$Q_s^{\lambda,\xi}$ is obtained from $Q_s$ with the following replacement for $j_{\rm min}<j<j_{\rm max}$:
\begin{eqnarray}
\Gamma^{v_j}_{r_j}
\rightarrow
\Gamma^{v_j}_{r_j}
{\rm e}^{i \, v_j \lambda_{r_j}},
\;\;\;\
\gamma(l_j)
\rightarrow
\gamma(l_j)+i \, \xi \, N_{l_j}.
\label{eqn:replacement1}
\end{eqnarray}
%
The first replacement was pointed out by Bagrets and Nazarov~\cite{Bagrets1,Qua-Noi-Mes-Phy}: The counting field for the current can be absorbed in the off-diagonal components.
Our result is the second expression: The counting field for the electron number can be absorbed in the diagonal components.

To proceed the calculation, the formal similarity between the master equation and the Schr\"odinger equation is utilized~\cite{Bagrets1,Qua-Noi-Mes-Phy}: 
$\partial_t \, |p(t) \rangle \!=\! -\hat{L} \, |p(t) \rangle$,
where 
$\langle n | p(t) \rangle=p_n(t)$
and 
$\langle n| \hat{L} | m \rangle=L_{nm}$. 
The operator $\hat{L}$ can be decomposed into three parts corresponding to Eqs.~(\ref{eqn:ODelement}) and (\ref{eqn:Delement}),
\begin{eqnarray}
\hat{L} = \hat{\gamma} - \sum_r (\, \hat{\Gamma}_r^{+} + \hat{\Gamma}_r^{-} \,).
\end{eqnarray}
Then, Eq.~(\ref{eqn:Dyson}) is rewritten as the Dyson series, 
\begin{eqnarray}
\cf(\lambda,\xi)
&=&
\!\!
\sum_m
\langle m|
\, 
\hat{T}
\exp
\!
\left (
-
\int_{-t_\infty}^{t_\infty}
\!\!\!\!
d t \, 
\hat{L}(t)
\right )
\!
|n_0 \rangle
\nonumber \\
&\approx&
\sum_m
\langle m|
\exp
\left (
-
t_0 \,
\hat{L}_{\lambda,\xi}
\right )
|\, p_0 \rangle,
\label{eqn:cgfmasterlongtime}
\end{eqnarray}
where $\hat{L}(t)=\hat{L}_{\lambda,\xi}$ for $|t| \!<\! t_0/2$ and $\hat{L}$ otherwise.
To obtain the second line, we utilized the assumption $t_0 \gg 1/\Gamma$
and the condition $\sum_m \langle m| \, \hat{L}=0$. 
We also assumed that the stationary state $|\, p_0 \rangle$, 
namely the zero eigen state $\hat{L} \, |\, p_0 \rangle=0$,
can be reached well before $t=-t_0/2$. 
The operator $\hat{L}_{\lambda,\xi}$ is obtained from $\hat{L}$ 
with the replacement (\ref{eqn:replacement1}):
\begin{equation}
\hat{L}_{\lambda,\xi}
=
\hat{\gamma}
+
i \, \xi
\sum_l N_l \, |\, l \rangle \langle l \, |
-
\sum_{r}
\left(
\hat{\Gamma}_{r}^+ \, {\rm e}^{i \, \lambda_{r}}
+
\hat{\Gamma}_{r}^- \, {\rm e}^{- \, i \, \lambda_{r}}
\right).
\label{eqn:LLX}
\end{equation}
In the limit of $t_0 \rightarrow \infty$, the minimum eigenvalue 
$\Lambda_{\rm min}$ of the operator $\hat{L}_{\lambda,\xi}$ is dominant
$
\cf
\! \approx \! 
\exp(-t_0 \, \Lambda_{\rm min})
$. 
Thus the CGF reads, 
\begin{eqnarray}
\cgf(\lambda,\xi)
=
-t_0 \, \Lambda_{\rm min}. 
\label{eqn:CGFmastereq}
\end{eqnarray}
%
It is noticed that the normalization condition 
$W(0,0)=0$
is satisfied since 
$\Lambda_{\rm min}$ 
is expected to be the zero eigen value for $\lambda_r=\xi=0$. 

%
%

Let us apply the above mentioned scheme to the RLM. 
The QD states are limited to the empty 
$|0 \rangle$
and singly occupied 
$|1 \rangle$
states. 
Then, 
$|p(t) \rangle$ is
given by 
$^t(p_1(t),p_0(t))$ 
and the operator~(\ref{eqn:LLX}) reads, 
\begin{eqnarray}
L_{\lambda,\xi}
 =
\left(
\begin{array}{cc}
\Gamma^- - i \xi/2 & -\Gamma^+(\lambda) \\
-\Gamma^-(\lambda) &  \Gamma^+ + i \xi/2
\end{array}
\right), 
\label{mat:LRLM}
\end{eqnarray}
\begin{eqnarray}\Gamma^\pm(\lambda)
=
\Gamma_{L}^\pm
{\rm e}^{\pm i \, \lambda_{L}}
+
\Gamma_{R}^\pm
{\rm e}^{\pm i \, \lambda_{R}}, 
\nonumber
\end{eqnarray}
in the matrix form. 
Here we shifted the origin
[$N_l$ in Eq.~(\ref{eqn:LLX}) is replaced with $N_l-1/2$].
As already demonstrated, unlike the counting field $\lambda$ in the off-diagonal components, the counting field $\xi$ appears in the diagonal components: $\mp\,i\, \xi/2$ for the occupied/empty state. 
Two eigenvalues of the matrix (\ref{mat:LRLM}) are 
\begin{eqnarray}
\Lambda_\pm
=
\Gamma_\Sigma(1 \pm \sqrt{D})/2.
\nonumber 
\end{eqnarray}
From $\Lambda_-$ and Eq.~(\ref{eqn:CGFmastereq}), we reproduce Eq.~(\ref{eqn:cgf1st}).

\subsection{Spin effect for $U \rightarrow \infty$}
\label{subsec:mastereq_result}

Let us discuss the spinful case in the presence of strong Coulomb interaction based on the scheme in the preceding section. 
The probabilities to find the available three states, 
$|\! \uparrow \rangle$, $|\! \downarrow \rangle$ and $|0 \rangle$
are given by 
$^t(p_\uparrow(t),p_\downarrow(t),p_0(t))$. 
The corresponding transition matrix 
with the counting fields reads, 
\begin{eqnarray}
L_{\lambda,\xi}
 =
\left(
\begin{array}{ccc}
\Gamma^- - i \xi/2 & 0 & -\Gamma^+(\lambda) \\
0 & \Gamma^- - i \xi/2 & -\Gamma^+(\lambda) \\
-\Gamma^-(\lambda) & -\Gamma^-(\lambda) & 2 \Gamma^+ + i \xi/2
\end{array}
\right). 
\label{eqn:transitionmatirx}
\end{eqnarray}
We found a unique eigen value $\Lambda_{\rm min}$ satisfying the normalization condition. 
The resulting CGF becomes Eq.~(\ref{eqn:cgf1st}) with $\Gamma_{r}^+$ replaced with $2 \Gamma_{r}^+$. 

Figure~\ref{fig:P_N} (b) shows the probability distribution of electron number in equilibrium. 
We observe the exponential distribution both 
for $\epsilon_0 \! \gg \! T$ and $-\epsilon_0 \! \gg \! T$ (lines E and C), but with different exponents by a factor 2:
In the former case, the excited state is $|\! \uparrow \rangle$ or $|\! \downarrow \rangle$. 
Therefore the decay process is the outgoing process of the spin inside the QD and the decay rate is $\Gamma_\Sigma \! \approx \! \Gamma$. 
In the latter case, the excited state is the empty state $|0 \rangle$. 
Thus the decay process is the incoming process of an up- or down-spin. 
Then the decay rate is the sum of up- and down-spin tunneling rates $\Gamma_\Sigma \! \approx \! 2 \Gamma$.

\section{Joint probability distribution of current and electron number}
\label{sec:3}

\begin{figure}[ht]
\includegraphics[width=0.85 \columnwidth]{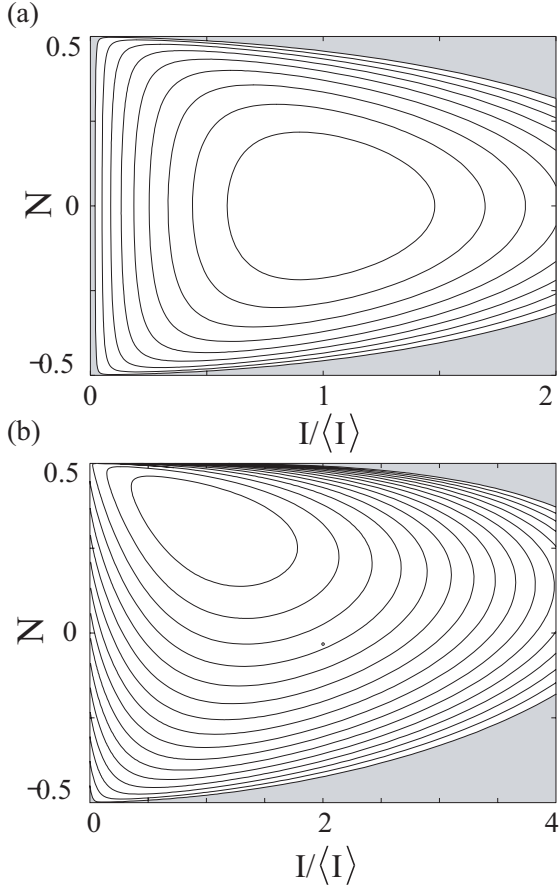}
\caption{
Contour plots of the joint probability distribution $\ln P(I,N)$ are given 
for the symmetric case in (a) and the asymmetric case in (b) ($\Gamma_{L} \!=\! 5 \Gamma_{R}$). 
The function $\ln P(I,N)$ has a maximum value $\ln P(I,N)=0$ at $I \!=\! \langle I \rangle$ and 
$N \!=\! \langle N \rangle \!=\! (\Gamma_{L} \!-\! \Gamma_{R})/2\Gamma$. 
The counter interval is $t_0 \Gamma/20$ and the shaded regions are for very small values. 
Vertical axes are shifted by $-1/2$.
}
\label{fig:contour}
\end{figure}

In this section, we discuss the joint probability distribution $P(I,N)$ 
in the limits of large source-drain voltage and weak tunneling 
($\mu_{L} \!=\! -\mu_{R}$ and $\epsilon_0 \!=\! 0$ for $\Gamma \! \ll \! eV$ at $T\!=\!0$). 
This case, the dot-level is inbetween the two chemical potentials, and Eq.~(\ref{eqn:cgf1st}) is also valid even at zero temperature. 
These conditions were realized in the experiment~\cite{Gustavsson}, 
but only the current distribution was measured.
Figures~\ref{fig:contour} (a) and (b) are contour plots of $\ln P(I,N)$. 
The inverse Fourier transform of the characteristic function $\cf$ is performed within the saddle point approximation, 
\begin{eqnarray}
\ln P(I,N)
\approx
\cgf^{(1)}
(\lambda^*,\xi^*)
-i \, t_0 I \, \lambda^*
-i \, t_0 N \, \xi^*,
\nonumber 
\\
i \, t_0 \, I \! = \! \partial_\lambda 
\cgf^{(1)}
(\lambda^*,\xi^*), 
\;\;\;\;
i \, t_0 \, N \! = \! \partial_\xi
\cgf^{(1)}
(\lambda^*,\xi^*).
\nonumber
\end{eqnarray}
%
For the symmetric case (a), we observe a clear feature: 
For $I \! \approx \! 0$, the distribution of $N$ is almost uniform, 
while for $I \! \gg \! \langle I \rangle$ a peak appears around $N=0$. 
The former corresponds to the rare tunneling case. 
In this limit, only one tunneling event occurs during $t_0$ and one kink appears in the experimental data [upper panel of Fig.~\ref{fig:1}]. 
Since the position of the kink is arbitrary, $\tau$ 
takes any value between $-t_0/2$ to $t_0/2$ with the same probability. 
The latter corresponds to the frequent tunneling case [lower panel], where many kinks are randomly distributed and thus 
$\tau \! \approx \! t_0 \langle N \rangle \!=\! 0$. 
We would like to stress that it is not obvious that the crossover between the two behaviors occurs at $I \!=\! \langle I \rangle$ and $N \!=\! \langle N \rangle$ as demonstrated in Fig.~\ref{fig:contour} (a). 
For the asymmetric case, the above described features are modified [Fig.~\ref{fig:contour} (b)]. 
Equidistance contours around $I\! \approx \!0$ implies 
that the electron number is exponentially distributed.
At the same time, a long tail appears in the regime $I \! > \! \langle I \rangle$, 
which implies that the current is Poisson distributed. 

\section{Effect of large quantum fluctuations}
\label{sec:4}

For now, we have mainly analyzed the number distributions and the joint probability distributions within the Markov approximation in Secs.~\ref{subsec:expdist}, ~\ref{sec:2} and \ref{sec:3}. 
In this section, we turn our attention to the non-Markovian effect enhanced by increasing strength of tunneling. 
Section~\ref{subsec:set} sketches the derivation of the CGF for the SET.
In order to go beyond the Markov approximation, the Keldysh generating functional approach is adopted again. 
Then in Sec.~\ref{subsec:rlmset}, we will compare the probability distributions of electron number for two cases, the RLM and the SET.

\subsection{SET in the intermediate conductance regime}
\label{subsec:set}

Here we consider the SET, the metallic QD. 
The energy levels inside the metallic island are continuous. 
However, the single-electron charging energy $E_C$ can be larger than the temperature and thus the low-energy physics is dominated by two island states, the charge neutral state $|0 \rangle$ and the state with an excess charge $e$ $|1 \rangle$. 
The energy difference between the two states 
$\Delta_0=E_C (1-2 Q_{\rm G}/e)$ is controlled by the gate-induced charge $Q_{\rm G}$ ($\Delta_0$ corresponds to $\epsilon_0$ for the RLM). 
The Hamiltonian is written with a pseudo-spin-1/2 operator acting on the two states 
$\hat{\sigma}_+ \! = \! |1 \rangle \langle 0|$ 
and 
$\hat{\sigma}_z \! = \! |1 \rangle \langle 1| - |0 \rangle \langle 0|$ 
as~\cite{Matveev}, 
\begin{eqnarray}
\hat{H}
&=& 
\sum_{r=L,R,I}
\sum_{k m}
\varepsilon_{{r} k} \,
\hat{a}_{{r} k m}^{\dag} \hat{a}_{{r} k m}
+ 
\Delta_0 \, \hat{\sigma}_z/2
\nonumber \\
&+& 
\sum_{r=L,R}
\sum_{k k' m}
(
T_{r} 
\hat{a}_{{\rm I} k m}^{\dag} 
\hat{a}_{{r} k' m}
\hat{\sigma}_{+}
+
{\rm H. c.}).
\label{eqn:Kondo}
\end{eqnarray}
The operator $\hat{a}_{{r} k m}^\dag$ creates an electron with the wave vector $k$ in the left or right electrode or the island (r=L,R,I). 
The transverse coupling describes the tunneling, which is assumed to preserve the spin and the transverse channel (The channels are denoted by $m$). 
The junction conductance is written with the tunneling matrix element $T_{r}$, the number of channels $N_{\rm ch}$ and the electron DOS $\rho_{r}$ as 
$1/R_{r} 
 =  2 \pi e^2 N_{\rm ch} |T_{r}|^2 \rho_{\rm I} \rho_{r}$. 
A measure for the tunneling strength is the dimensionless conductance 
$\alpha_0=\alpha_0^{L}+\alpha_0^{R}$, 
where 
$\alpha_0^{r}=1/2 \pi e^2 R_{r}$~\cite{SchoellerSchoen}.

The source fields for current and number, $\varphi_{r}$ and $h$, are introduced following Ref.~\cite{Utsumi}:
\begin{eqnarray}
T_{r} \! \rightarrow \! T_{r} e^{i \varphi_{r}(t)},
\;\;\;\;
\Delta_0 \! \rightarrow  \! \Delta_0-h(t). 
\end{eqnarray}
They are fixed in the same way as for the RLM case, 
Eqs.~(\ref{eqn:so_ph}) and (\ref{eqn:so_qu}). 
The voltage drop in the junction $r$, $\mu_r$, is determined by the capacitances of the left and right junctions $C_L$ and $C_R$ as $\mu_{L(R)}=\kappa_{L(R)} eV$ [$\kappa_{L(R)}=\pm C_{R/L}/(C_L+C_R)$].

\begin{figure}[ht]
\includegraphics[width=0.9 \columnwidth]{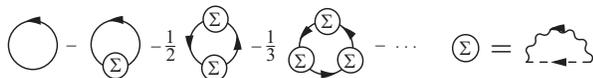}
\caption{
Diagrammatic expansion of $\cgf$. 
Solid, dashed and wavy lines are for Dirac, Majorana and particle-hole Green functions. 
}
\label{fig:diagram}
\end{figure}

The following calculations are completely in parallel to those in Ref.~\cite{Utsumi2}, therefore we will just briefly sketch them. 
The starting point is the effective action of the pseudo-spin for $N_{\rm ch} \! \rightarrow \! \infty$ derived from the Hamiltonian~(\ref{eqn:Kondo}) after tracing out the electron degrees of freedom. 
By introducing the Majorana representation~\cite{Spencer},
$\hat{\sigma}_z=2 \hat{c}^\dagger \hat{c}-1$ 
and 
$\hat{\sigma}_+=\hat{c}^\dagger \hat{\phi}$, 
where $\hat{c}$ and $\hat{\phi}$ are Dirac and Majorana fermions, the parts describing the charging energy $S_{\rm ch}$ and the tunneling $S_t$ are written as~\cite{Utsumi,Utsumi1,Utsumi2},
\begin{eqnarray}
S_{\rm ch}
&=& 
\int _C \! d t 
\left \{ 
c(t)^* 
\left [
i \partial_t -\! \Delta_0+h(t) 
\right ] 
c(t) 
+ 
\frac{\ri}{2} \phi(t) \partial_t \phi(t) 
\right \},
\nonumber \\
S_t
&=&
-
\int _C \!\! d t \, d t' \,
c^*(t) \, \phi(t) \, \alpha (t,t') \, \phi(t') \, c(t'), 
\nonumber
\end{eqnarray}
where $c$ and $\phi$ are the complex and the real Grassmann fields, 
respectively. 
It is the Bose Kondo model with anisotropic coupling and the magnetic field. 
The bosonic propagator, the particle-hole GF 
$
\alpha (t,t')
=
\sum_{r=L,R}
\alpha_r (t,t')
{\rm e}^{i(\varphi_r(t)-\varphi_r(t'))}
$
describes the tunneling of an electron from the lead $r$ into the island. 
The bare part is Ohmic except for the offset and written in the Keldysh space as,
\begin{eqnarray}
\tilde{\alpha}_r(\omega)
\!=\!
-i \pi \, \alpha_0^r \, 
\frac{(\omega \!-\! \mu_r) \, E_C^2}
{(\omega \!-\! \mu_r)^2 + E_C^2}
\left(
\begin{array}{cc}
0 & -1 \\
1 & \displaystyle 
2 \coth \frac{\omega-\mu_r}{2T}
\end{array}
\right).
\nonumber
\end{eqnarray}
The high-energy cutoff $E_C$ is introduced to regularize the UV divergence. 
The path-integral is performed by introducing a Grassmann source field $J$, 
\begin{eqnarray}
\cf
&=&
\int \!\! D [c^*,c,\phi] \,
\exp (\, i \, S_{\rm ch}+i \, S_t)
\nonumber \\
&=&
\exp \! 
\left(-\sum_n \frac{(-1)^{n}}{n} 
{\rm Tr} \left[ \left( g_c 
\frac{\delta}{\delta J} 
\alpha
\frac{\delta}{\delta J} 
\right)^n \right] \right) 
\nonumber \\ 
&\times& 
\left. 
\exp \left( \frac{i}{2} \int_C \rd 1 \rd 2 \, J(1) g_{\phi}(1,2) J(2) \right) \right|_{J=0} 
{\rm e}^{{\rm Tr} \left[ \ln g_c^{-1} \right]}.
\nonumber 
\end{eqnarray}
The trace is understood as time integration over $C$. 
The Dirac and Majorana GFs, 
$g_c$ and $g_\phi$, 
satisfy, 
\begin{eqnarray}
\left [
i \partial_t -\! \Delta_0+h(t) 
\right ] \, 
g_c(t,t')
&=&
\delta(t,t'), 
\nonumber
\\
i \partial_t \, g_\phi(t,t')
&=&
\delta(t,t'), 
\nonumber
\end{eqnarray}
and the anti-periodic boundary condition such as Eq.~(\ref{eqn:antiperiodic}). 
Further calculations are based on the diagrammatic expansion~\cite{Utsumi}. 
The diagrams in Fig.~\ref{fig:diagram} are resummed in our approximation~\cite{Utsumi1,Utsumi2,Utsumi}.
\begin{eqnarray}
\cgf^{\rm RTA}={\rm Tr}
\ln \left( i G_c^{-1} \right),
\;\;\;\;
G_c^{-1}
=
g_c^{-1}
-
\Sigma,
\label{eqn:cgfrta}
\\
\Sigma(t,t')
=
-i \, g_\phi(t',t) \, \alpha (t,t').
\end{eqnarray}
We can see Eq.~(\ref{eqn:cgfrta}) is formally the same as Eq.~(\ref{eqn:CGFLL}).
Then it is almost clear that the result is formally the same as Eq.~(\ref{eqn:CGF}), but the Keldysh and retarded components of the self-energy corresponding to Eqs.~(\ref{eqn:reseenrlm}) and (\ref{eqn:keseenrlm}) are replaced with those of the SET: 
\begin{eqnarray}
\Sigma_r^R(\omega)
&=& 
2 \, \alpha_0^r
\left[
{\rm Re} \, \psi \! \left( i
\frac{\omega-\mu_r}{2 \pi T} \right)
-
\psi \! \left( \frac{E_C}{2 \pi T} \right)
- \frac{\pi T}{E_C}
\right]
\nonumber \\
&-& 
i \pi \, 
\alpha_0^r \,
(\omega-\mu_r)
\coth \! \left( \frac{\omega-\mu_r}{2 \, T} \right), 
\nonumber
\\
\Sigma_r^K(\omega)
&=& 
-2 i \pi \, \alpha_0^r \, (\omega-\mu_r),
\nonumber
\end{eqnarray}
where $\psi$ is the digamma function. 
Consequently, the transmission probability Eq.~(\ref{eqn:T}) reads, 
\begin{eqnarray}
{\cal T}(\omega)
= 
\frac{(2 \pi)^2
\prod_{r=L,R}
\alpha_0^r
(\omega-\mu_r)
\coth 
\left( 
\displaystyle
\frac{\omega-\mu_r}{2 T} 
\right) 
}
{|\omega - \Delta_0 - 
\sum_{r=L,R} 
\Sigma^R_r(\omega)|^2}. 
\nonumber
\end{eqnarray}
It reproduces the transmission probability derived in Ref.~\cite{SchoellerSchoen}, which can describe the higher order inelastic cotunneling~\cite{Nazarov} as well as the sequential tunneling.

\subsection{Probability distributions of electron number for SET and RLM}
\label{subsec:rlmset}

In the limit of $\alpha_0 \! \rightarrow \! 0$, the CGF $\cgf^{\rm RTA}$ reproduces that of the extended master equation approach. 
The formal expression is the same as Eq.~(\ref{eqn:cgf1st}) with the tunneling rate of the RLM, $\Gamma_r^\pm$, replaced with that of the SET, 
\begin{eqnarray}
\Gamma_{rI/Ir}  = \pm 
\frac{\Delta_0 \!-\! \mu_{r}}
{e^2   R_{r} \, ({\rm e}^{\pm(\Delta_0 - \mu_{r})/T} \!-\! 1)}.
\nonumber 
\end{eqnarray}
Therefore within the Markov approximation, though this time the tunneling rate depends on the source-drain voltage, the current and number obey qualitatively the same statistical distribution as that of the RLM. 

%
%

\begin{figure}[ht]
\includegraphics[width=0.85 \columnwidth]{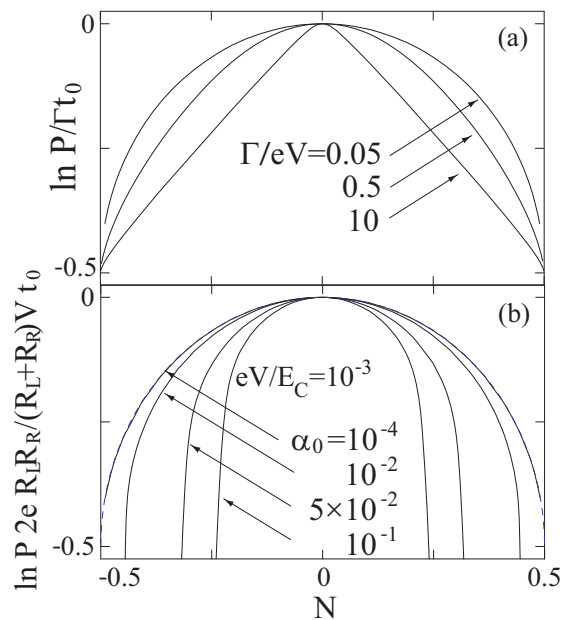}
\caption{
The nonequilibrium number distributions are given for the resonant-level model in (a) and the single-electron transistor in (b) at $\epsilon_0 \!=\! \Delta_0 \!=\! 0$ and $T\!=\!0$. 
Solid lines are for various coupling strengths $\Gamma$ in (a) and for various dimensionless tunnel conductances $\alpha_0(=\sum_{r=L,R} 1/2 \pi e^2 R_{r})$ in (b). 
The dashed lines 
overlapping with curves for $\Gamma/eV \!=\! 0.05$ (a) and for $\alpha_0 \!=\! 10^{-4}$ (b) are obtained by the extension of the master equation approach. 
Horizontal axes are shifted by $-1/2$. 
}
\label{fig:P_NN}
\end{figure}

However, it is not the case when the tunnel coupling increases and the quantum fluctuations are enhanced. 
We compare the nonequilibrium number distributions for the RLM [Fig.~\ref{fig:P_NN} (a)] and for the SET [Fig.~\ref{fig:P_NN} (b)] at $\epsilon_0 \!=\! \Delta_0 \!=\! 0$ in the symmetric case ($\Gamma_{L} \!=\! \Gamma_{R}$, $\alpha_{L}^0 \!=\! \alpha_{R}^0$, and $\mu_{L} \!=\! -\mu_{R}$). 
At very small tunnel coupling, the distributions for both cases approach the master equation results [dashed lines overlapping with curves for $\Gamma/eV \!=\! 0.05$ in panel (a) and for $\alpha_0 \!=\! 10^{-4}$ in panel (b)].
Here, vertical axes are normalized by $\Gamma_\Sigma$ at 0K, namely 
$\Gamma$ for the RLM and 
$(R_{L}+R_{R}) V/2 \, e R_{L} R_{R}$ for the SET. 
For the large tunnel coupling ($\Gamma/eV$ and $\alpha_0$), 
the fluctuations of the electron number get stronger and thus the distributions for both cases shrink. 
However, a qualitative difference appears: 
For the SET, the width of the distribution just shrinks, but for the RLM, 
a two sided exponential distribution like profile shows up. 
We can expect that it is related with the difference in the dominant higher order tunneling processes. 
For the RLM, the large tunnel coupling enhances the multiple tunneling process of a single electron, which preserves the phase coherence. 
For the SET with large $N_{\rm ch}$, the coherent process, the elastic cotunneling process, is suppressed as compared with the inelastic cotunneling process~\cite{SchoellerSchoen}. 
In the latter process, different electrons tunnel back and forth between the leads and the metallic island. Therefore, the phase coherence is lost.

The phase coherence effect is pronounced when the QD conductance 
approaches to the quantum conductance $e^2/(2 \pi \hbar)$. 
In this regime, the high band-width of the QPC electrometer is required -- even for a small source-drain voltage, 0.1 mV, the band-width is required to be over than 10 GHz. 
For accurate measurements, the QPC current should be much larger than 1 nA, which would make the back action neglected here a crucial ingredient~\cite{Nazarov}. 
The FCS theory of electron number including the QPC electrometer will be discussed elsewhere.

\section{Summary}
\label{sec:5}

%
%

In conclusion, we have extended the FCS theory for the statistical distribution of electron number inside the QD. 
We found the non-Gaussian distribution, the exponential distribution, when the dot-level is far away from the lead chemical potentials. 
The exponent is proportional to the spin degeneracy. 
The large tunnel coupling would narrow the number distribution generally. 
Especially for the symmetric noninteracting QD, the two sided exponential distribution appears in the probability distribution of electron number out of equilibrium. 
One measurable prediction for the presently available experimental setups is the joint probability distribution of current and electron number, which reveals the nontrivial correlations between the two quantities. 

%
%

In the present paper, we have extended the FCS theory of the master equation approach~\cite{Bagrets1,Qua-Noi-Mes-Phy} and derived the general expression for the CGF of the joint probability distribution. 
We have shown that the CGF is obtained from the minimum eigenvalue of the transition matrix with the counting field for electron number in the diagonal components (Sec.~\ref{subsec:mastereq}). 
The non-Markovian effects have been treated in the frame of the Keldysh generating functional for the two cases, the non-interacting (RLM) and the interacting (SET) QDs. 
For the RLM, it is possible to derive the exact expression (Sec.~\ref{subsec:rlm}). 
For the SET, we performed resummation of diagrams~\cite{Utsumi,Utsumi1,Utsumi2} to obtain an approximate CGF in the intermediate conductance regime (Sec.~\ref{subsec:set}).

\begin{acknowledgements}

I thank D. Bagrets, T. Fujisawa, A. Furusaki, Y. Gefen, 
T. Hayashi, T. L\"ofwander, T. Martin and G. Sch\"on for valuable discussions. 
Especially, I thank D.~Bagrets for the in-depth discussions on the extension of the master equation approach (Sec.~\ref{subsec:mastereq}). 
This work was supported by 
RIKEN Special Postdoctoral Researchers Program
and 
the DFG-Forschungszentrum \lq\lq Centre for Functional Nanostructures". 

\end{acknowledgements}

\appendix

\end{document}